# Fake Plastic Voters:

## When Political Parties Can Use AI-Simulated Focus Groups

Claudio Novelli[1], Javier Argota Sánchez-Vaquerizo[2], Jennifer Cyr[3], Giuliano Formisano[4], Simon McDougall[1], Giulia Sandri[5], Luciano Floridi[1,6]

Corresponding author: claudio.novelli@yale.edu

**Abstract.** Political parties strive to understand their electorates, and focus groups are a vital tool in these efforts. AI-enhanced simulation technologies (AESTs) enable synthetic focus groups in a fraction of the time (and cost), raising the question of when and how such simulated evidence can be used in campaign research. This paper develops a decision matrix to help party strategists match research needs to appropriate simulation technologies and to identify when to escalate to hybrid or fully human focus groups. The matrix combines three dimensions: strategic purpose, deployment risk, and empirical grounding of the simulation tool. Strategic purpose is the decisive dimension, as it determines what kind of evidence the focus group is meant to produce: observing how political meanings and identities emerge through interaction (Mode 1) or testing and refining campaign messages (Mode 2). The matrix shows that, given documented failure modes such as sycophancy, persona drift, and the suppression of minority viewpoints, AESTs cannot replace human interaction in Mode 1 at any risk level. Within Mode 2, suitability depends instead on deployment risk and on the empirical grounding. Yet even here, we caution that routine reliance on AESTs may erode the qualitative craft on which sound judgment depends.

## 1. Introduction

Researchers are increasingly turning to AI-based simulations to create digital settings in which social interactions can be generated and studied (Argyle et al. 2023; Park et al. 2023; Horton 2023). One potential real-world outlet for these simulations is in campaign strategy. Political parties regularly use focus groups to understand their electorates. AI-enhanced simulation technologies (AESTs) could revolutionize this strategy, transforming wholly human-based conversations into one of several different focus group formats at a party's disposal. The early record of these technologies is mixed. Simulated respondents have reproduced aggregate human response patterns with some accuracy on certain tasks, at a fraction of the cost and time of fieldwork (Argyle et al. 2023; Horton 2023), but replication studies also document homogenized outputs, demographic caricature, and estimates that diverge from — or even reverse — those obtained from real participants (Santurkar et al. 2023; Bisbee et al. 2024). Many warnings have already been moved against this approach from a more theoretical point of view (Dillion et al. 2023; Li and Tao 2026). Therefore, our concern is: when do these tools offer real value, and when do they instead create a misleading sense of certainty?

[1] Digital Ethics Center, Yale University, New Haven, CT, USA
[2] Department of Humanities, Social and Political Sciences (GESS), ETH Zürich, Switzerland
[3] Department of Political Science and International Studies, Universidad Torcuato Di Tella, Buenos Aires, Argentina
[4] Department of Political Science, University of Zurich, Zurich, Switzerland
[5] CEVIPOL, Université Libre de Bruxelles, Brussels, Belgium
[6] Department of Legal Studies, University of Bologna, Bologna, Italy

We answer this question by proposing a decision matrix that helps party strategists and campaign practitioners match research needs to appropriate simulation technologies, and advises when to escalate to hybrid validation or fully human focus groups. It does so by combining three dimensions: the strategic purpose of the focus group, the deployment risk, and the degree of empirical grounding in the simulation technique. The matrix does not identify "admissible cases" for automation. Rather, it delineates where simulation offers bounded, auxiliary utility before mandatory escalation to human validation.

The first dimension is the most important, because it determines what kind of evidence the focus group is meant to produce. Here, we distinguish between two strategic modes of focus-group use in party work. The first is Interaction-Based Meaning-Making (Mode 1), where the purpose is to observe how political meanings, identities, and justifications emerge through live group interaction. The second is Instrumental Decision-Support (Mode 2), where the purpose is to test messages, frames, and strategic options quickly and comparatively. This distinction functions as the matrix's first-order filter.

Let us illustrate these two modes with some examples. Operating in Mode 1, a party might convene undecided voters to observe how they reason together about immigration and, more specifically, which justifications gain traction or how positions shift as participants react to one another. Operating in Mode 2, the same party might need to know, within days, which of three slogans for a childcare policy performs best among suburban swing voters. In this case, the focus group works as a bounded, comparative task in which the group functions chiefly as a rapid testing device. Early applications of AESTs, from LLM-driven message pretesting to fully synthetic discussion panels, cluster around tasks of this second kind (Park et al. 2023; Zhang et al. 2024).

We argue that the two modes place very different demands on simulation. Mode 1 depends on emergent social dynamics that current AESTs cannot reliably reproduce. Documented failure modes such as sycophancy, persona drift, and the suppression of minority viewpoints make simulations poorly suited to capturing the open-ended interactional processes through which political meaning is negotiated in real groups. The reasons are structural. Since LLM training privileges mainstream language, and post-training alignment steers outputs toward agreeable defaults, simulated agents tend to converge on an "average" voice, under-representing dissent and minority positions even when explicitly prompted to voice them (Taubenfeld et al. 2024; Santurkar et al. 2023); Section 4 reviews this evidence in detail. Mode 2, by contrast, is better suited to simulation because it involves bounded, comparative tasks such as variant triage, scenario stress testing, and sensitivity analysis. In these cases, AEST outputs can be used more cautiously as inputs for robustness checks rather than treated as evidence in their own right.

The second dimension is deployment risk, that is, how consequential it would be if the simulation output turned out to be wrong or misleading. The final dimension is the empirical grounding of the AEST: how firmly the simulation is anchored in real data about actual voters, ranging from generic LLM personas at one end to fully human focus groups with AI augmentation at the other. By LLM personas, we mean synthetic discussion participants created by conditioning a large language model on a profile of demographic and attitudinal attributes, typically a natural-language description supplied in the prompt (Argyle et al. 2023; Park et al. 2023). By AI augmentation, instead, we mean configurations in which participants and moderation remain human and AI is confined to support tasks such as transcription and live summarization (Sections 3.3–3.4 detail these configurations and how persona attributes are operationalized).

Strategic mode shapes how the second and third dimensions apply. For Mode 1 tasks, the matrix returns the same answer regardless of risk or grounding level:

human-led focus groups are always required, and no simulation can substitute. It is only within Mode 2 that risk and grounding become the operative variables, producing concrete guidance: lower-grounded tools are acceptable only in low-stakes, internal settings (e.g., narrowing a long list of slogan variants before any human testing, or piloting a discussion guide); as risk increases, grounding requirements rise accordingly; and at the highest stakes (e.g., authorizing paid media buys, finalizing a manifesto commitment, or making public claims about voter opinion), simulations may inform what to test but should not authorize what to deploy without live human validation.

A theme that runs through the entire analysis is that the value of focus groups has never resided in their outputs alone, but in the human agency exercised in observing and interpreting them: the craft of reading what a group of citizens is actually doing. Our argument is accordingly not that simulation is useless, but that it must be staged so that this interpretive agency is preserved rather than displaced.

Our standpoint is that of researchers of computational social science, political science, and digital ethics who are normatively committed to democratic deliberation and cautious about automating high-stakes campaign decisions. This standpoint motivates our emphasis on the limits of AESTs in Mode 1 settings and on preserving human interpretive agency even when simulations are used in Mode 2 tasks. It also informs our choice to foreground failure modes that suppress minority viewpoints and homogenise political meaning, rather than treating simulated outputs as neutral reflections of "the electorate".

## 2. Focus groups in party politics: origins and weaknesses

A focus group is a purposively convened, typically small, group of people brought together by an observer, such as a social science researcher, to analyze their reactions and opinions on a predefined topic in a controlled environment (Morgan 1996, 130).

Robert Merton and his colleagues developed the "focused interview" at Columbia's Bureau of Applied Social Research in the 1940s precisely because the era's dominant quantitative instruments could record what audiences felt about wartime radio propaganda, but not why; only guided group discussion could surface the reasoning, associations, and misunderstandings behind measured responses (Merton and Kendall 1946). The technique then migrated into commercial market research, where it remains among the most widely used tools for product and message testing — and where AI-simulated respondents ("silicon samples") are already being adopted at scale (Sarstedt et al. 2024), anticipating the political developments examined in this paper. This break from the dominant quantitative paradigm frames the discussion that follows.

The use of focus groups in politics developed as part of the broader shift toward marketing-oriented campaigning (Wring 2005; Temple 2010). Parties use them — or are advised to do so by organizations with field experience (Canavor 2006) — to generate evidence of citizens' concerns, priorities, and the reasoning behind them. They not only elicit what people think, but also how they come to think in interaction: participants assess arguments, adopt or reject shared language, and negotiate what counts as "common sense" in real time. This is what gives the method its value for party politics.

However, as shown in the UK Labour Party's experience (Wring 2007, 77), such findings are not always used neutrally but may be selectively deployed or leaked to steer internal debates in favor of factional interests. This selective deployment exposes that evidence in party politics enters a contested internal arena in which factions and party leaders compete for authority (Scarrow et al. 2017). The temptation to commission multiple simulations until a preferred result appears, a form of "results shopping", could emerge as part of intra-party dynamics. Proposals that threaten established power centers are likely to be selectively diluted or shelved regardless of technical merit. When analyzing the adoption of an internal innovation, such as focus groups and AEST, it therefore matters who commissions them, who has access to raw outputs, and who arbitrates between competing findings.

Before delving into the political opportunity discussion, we should stress that focus groups offer several advantages over alternative tools such as surveys or questionnaires.[7] Surveys offer the benefit of statistical reliability, but they tend to be rigid: questions are fixed in advance, they can be misunderstood without the opportunity to clarify, and they capture answers in isolation rather than in conversation. Additionally, focus groups offer distinctive sampling advantages, such as encouraging contributions from people who feel they have little to say about a topic until they engage in discussion (Kitzinger 1995, 300).

[7] In practice, the two methods often work together. Researchers frequently run focus groups first to spot unclear wording or unexpected ideas, which then helps them shape better survey questions (Nassar-McMillan and Borders 2002).

Against this background, it emerges how much focus groups are conversation inducers. The group format lowers the threshold for discussing sensitive or stigmatized topics, and the interactive setting makes the method especially productive for generating new themes (Cyr 2019; Baden et al. 2022; Chen et al. 2025; Tracy 2010).

At the same time, focus groups pose practical, methodological, and ethical challenges. The analysis that follows focuses on those challenges most likely to arise in the study of party politics.[8]

### 2.1. Dominant voices, Groupthink, and Social Pressure

A key strength of focus groups lies in their ability to elicit how individuals articulate and negotiate their views through group interaction. Yet the same social dynamics that generate these insights can also distort the data. A common concern is that one or two dominant individuals may steer the discussion. Participants who are aggressive, outspoken, or accorded greater authority within the group - such as parents in a focus group on paternity-leave policy (Smithson 2000, 107 ff.) - can monopolize the conversation, marginalizing quieter voices. The resulting data may therefore capture the perspectives of the most assertive participants rather than a balanced range of views.

Relatedly, peer pressure and group norms can produce conformity. Participants often experience subtle social pressure to conform to the majority or to refrain from expressing views that might be disapproved of. This tendency toward socially desirable responding is well-documented in qualitative research (Bispo Júnior 2022; Smithson 2000) and is exacerbated when discussing political opinions that are often shaped by perceived norms, partisan identities, and expectations about how one should behave (Cohen 2003).

Politicians refer colloquially to *shy voters* — those who conceal support for a party in public but vote for it in private. The classic illustration is the "shy Tory" effect of the 1992 British general election, when polls and group discussions alike understated Conservative support because respondents were reluctant to admit it (Jowell et al. 1993). The opposite pattern also emerges: rather than moderating one another, relatively homogeneous groups can reinforce shared assumptions and escalate toward more extreme stances (Sagoe 2012).

[8] Other challenges to focus groups, beyond those discussed here, include lack of anonymity (reducing candor and participation), increased cognitive burden for technical questions, and interpretive difficulties posed by unstructured discussions and the volume of qualitative data.

These group dynamics are not inherently deceptive; they reflect how social interactions often manifest in real life. The difficulty is interpretive: capturing these interactional processes with precision and relating them to individual-level dispositions requires careful design and analysis. Otherwise, focus groups can drift from being tools of in-depth inquiry into superficial exercises that neglect the interactional texture that gives those statements meaning (Rook 2003; Savigny 2007).

Measures to address social desirability bias come with significant trade-offs. Researchers can ask participants to write down initial answers before discussion, ask the same question in multiple ways (Cyr 2019, 34–35), or assemble more homogeneous groups to make timid participants comfortable speaking (Smithson 2000, 108–9). These are sensible interventions, but focus groups are typically time-limited (one and a half to two hours), so extra steps reduce the number of topics that can be covered, and group-level data on the consensus-building process itself are obscured (Cyr 2019, 35).

These limitations are particularly acute in the context of party politics research. Political focus groups already face tight participation constraints: a 1.5–2-hour session can deter attendance in itself (Boas 2024), leaving little room for additional methodological steps. Political opinions are often shaped by perceived norms, partisan identities, and social desirability bias (Cohen 2003).

### 2.2. Replicability and Generalizability

We should clarify that, within the qualitative paradigm, replication and statistical generalization are not the method's objectives. Focus groups aim more at credibility, thick description, and transferability rather than repetition and extrapolation, and judging them by quantitative standards is, for the most part, a category error. What follows should therefore be read not as an indictment of the method on foreign criteria, but as an analysis of a tension specific to party politics: political actors routinely extract poll-like, quasi-quantitative conclusions from focus groups (Savigny 2007; Boas 2024), and it is this instrumental use that generates the replicability and generalizability pressures discussed below.

Replicability refers to the ability to repeat a study's procedures and obtain similar empirical patterns or broadly comparable analytic inferences. Methodologically, it signals a degree of systematicity in the design and implementation of a particular method.

In quantitative research, high replicability is a key contributor to reliability. In qualitative focus-group work, however, replicability is inherently difficult to achieve because the interactional dynamics of any given group cannot be reproduced exactly (Chioncel et al. 2003). For instance, an election-focused focus group conducted during one campaign cycle cannot be replicated precisely in a later cycle when candidates, issues, and public mood have shifted.

Scholars have proposed ways to enhance replicability at the design level, particularly through detailed documentation of recruitment, moderation, and protocol implementation (Cyr 2019; Boas 2024). Yet these measures have limits. Transparency can collapse into box-ticking that leaves interactional dynamics undocumented, especially in party-political settings where actors have incentives to shape what is recorded. Political contexts change rapidly: two teams running an identical guide on how to respond before and after a major scandal may elicit entirely different discussions. The procedural method is replicated, but the data-generating environment is not (Winters and Carvalho 2013).

Connected to the problem of replication, and shaped by group dynamics, is the issue of the generalizability of focus-group findings (Cyr 2019, 33), that is, the extent to which results can be extended beyond the immediate sample. Concerns about generalizability also matter for avoiding the normative risks highlighted by Savigny (2007), who argues that political-marketing uses of focus groups often overclaim "scientific" authority and narrow democratic debate by outsourcing political judgment to a small, highly selected set of voices.

Yet focus groups rely on small, non-random samples, typically six to ten individuals, selected purposively or through convenience methods rather than random sampling (Sagoe 2012, 8). Researchers often assemble relatively homogeneous groups to promote comfort and open discussion (Van Ingelgom 2020; Hennink 2014), partly to address the group-dynamics problems discussed above. Recruitment also tends to attract individuals with flexible schedules — e.g., retirees, students, underemployed workers — and paid participants[9], so that focus-group participants may differ systematically from the average voter (Merton 1987). Virtual focus groups ease some of these constraints but do not eliminate self-selection and may compound the professional-participant problem.

[9] Those participants who join focus groups for economic compensation.

Although statistical generalization is not typically a primary goal of focus groups, in political science, they are frequently used to inform inferences about a broader study population, and scholars (as well as practitioners) may be tempted to treat their findings as indicative of the broader population beyond the immediate sample (Boas 2024, 385). Against this background, the literature has explored ways to strengthen generalizability: random sampling combined with segmented homogeneous groups (Krueger 1988), using focus groups for hypothesis generation rather than direct generalization (Vicsek 2010, 126), and triangulation with quantitative data (Santos et al. 2020; Cyr 2017).

Each of these strategies, however, entails its own limitations. In random sampling, as discussed in the previous section, it can disrupt group dynamics and lead to unbalanced discussions (e.g., dominant voices), distorting the findings. Exploratory use risks transferring the biases of a small, self-selected group into the design of subsequent quantitative instruments, while triangulation can potentially obscure minority views and interactional dynamics central to political talk.

### 2.3. Logistics, Iteration, and Mistakes

Focus groups require substantial planning and logistical coordination. In-person sessions require participants and moderators to meet in the same place at the same time. Practical constraints (e.g., committing one to two hours) also shape the discussion itself by limiting the number of questions that can reasonably be addressed within the available time.

For political parties, these constraints have clear consequences. Individuals who are busy, highly mobile, or geographically dispersed are less likely to take part, skewing the findings. Virtual focus groups partially fix the access issue, but not the time-commitment one. Electoral cycles intensify the problem, often forcing researchers to run shorter sessions or fewer groups than would be methodologically ideal (Baden et al. 2022). Even when focus groups save time relative to one-on-one interviews by capturing multiple opinions in a single session, they still require lead time for recruitment and careful moderation. Compounding this, focus-group research is typically *iterative*: methodological guidance recommends several groups per target demographic to capture a range of perspectives, identify patterns, isolate outliers, and reach thematic saturation.

Iteration also shapes design and analysis. After an initial session, researchers typically refine their discussion guide and moderation techniques. Baden et al. (2022) propose "serial focus groups" as an explicitly longitudinal design in which the same participants meet several times, allowing researchers to observe how opinions and dynamics develop. Political campaigns rarely reconvene the same individuals, but they run successive groups across regions or campaign stages to test and refine messages, making the research cyclical by design.

The iterative approach, in whatever form, may improve the robustness of findings but also requires additional time and coordination, a trade-off that practitioners must balance against tight campaign timelines. Online focus groups can mitigate logistical costs by making it easier to recruit participants who would otherwise be hard to reach (Stewart and Shamdasani 2017). Yet they do not eliminate the time required for substantive discussions and meaningful iteration.

Iteration can also be driven by necessity rather than design, as organizing a focus group presents numerous opportunities for error. When mistakes occur, they frequently require the session to be repeated or redesigned, steps that can be difficult to schedule or simply infeasible. In political contexts, delays caused by re-running groups may undermine a party's ability to shape its strategy around emerging findings.

We can imagine at least four types of issues in this context. The first involves fieldwork logistics: participants may fail to show up, or unexpected problems may arise with the venue. A second concern is recruitment: researchers may discover too late that the sample is inappropriate, perhaps including too few participants, participants that do not adequately represent the target population, or individuals unwilling to engage with one another. The third one relates to question design: poorly phrased questions (whether too complex or unclear) can produce unusable data. The fourth type of mistake concerns moderation: political parties and research teams often delegate focus-group facilitation to assistants or junior staff, which can intensify observer effects when moderators lack training or confidence.

In practice, when early focus groups suffer from these problems, researchers may treat them as pilot sessions and adjust recruitment strategies, question guides, or facilitation techniques in subsequent groups (a point we will return to later). This controlled form of re-running is one way to bolster research quality, but it also exemplifies why iteration is both unavoidable and resource-intensive.

## 3. AI-Enhanced Simulation Technologies (AEST) for Focus Groups

The political application examined here follows, rather than leads, a commercial trajectory, retracing the path of the focus group itself from Merton's Bureau to the marketing industry. In contemporary market research, industry surveys report that a large majority of researchers have already used synthetic responses, and major providers now offer synthetic panels and 'digital twin' audiences as standard products (Qualtrics 2024; Kalidas and Ostler 2024). The evidence on efficacy is instructive for our argument, as LLM-simulated consumers have been shown to approximate human willingness-to-pay and concept-test results at a fraction of the cost, but the accuracy is uneven, improving with grounding in prior survey data, and is strongest for bounded comparative tasks (Sarstedt et al. 2024; Brand et al. 2023).

While direct applications of AESTs to the focus-group format remain emergent, the related literature is no longer limited to isolated proofs of concept. Recent contributions span fully synthetic sessions and AI-supported moderation with human participants (Zhang et al. 2024), and assess how AI support shapes engagement and analytical efficiency in virtual focus groups (Chen et al. 2025). Complementarily, methodological considerations are also emerging (Filipova et al. 2025).[10] A 2023 report by the European Commission's Joint Research Centre shows how LLM-driven agents can serve as believable proxies for human behavior in simulated environments (Hradec et al. 2023).

Several techniques can be employed to simulate focus groups, drawing on (theoretical and empirical) research on human group simulations more broadly. In turn, these techniques build on decades of (pre-AI) simulation work, such as agent-based modeling and system dynamics, now increasingly augmented by AI-powered components.

Rather than forming mutually exclusive modeling approaches, simulated focus groups are best understood as a design space defined by two orthogonal features (see Figure 1): (1) how simulated participants are constructed and empirically grounded, and (2) how the discussion is deployed. A cross-cutting design choice concerns the behavioral engine used to generate participant behavior (e.g., rule-based heuristics, agent-based models, LLMs, or hybrids). We foreground grounding and deployment because they determine what kind of evidence a simulation can plausibly support, while the engine choice primarily affects realism, variance, and failure modes.

[10] An earlier study explored whether three-dimensional virtual worlds (specifically Second Life) could function as a valid setting for focus group research. In that work, however, the avatars were digital stand-ins controlled by real human participants, not synthetic individuals (Gadalla et al. 2016).

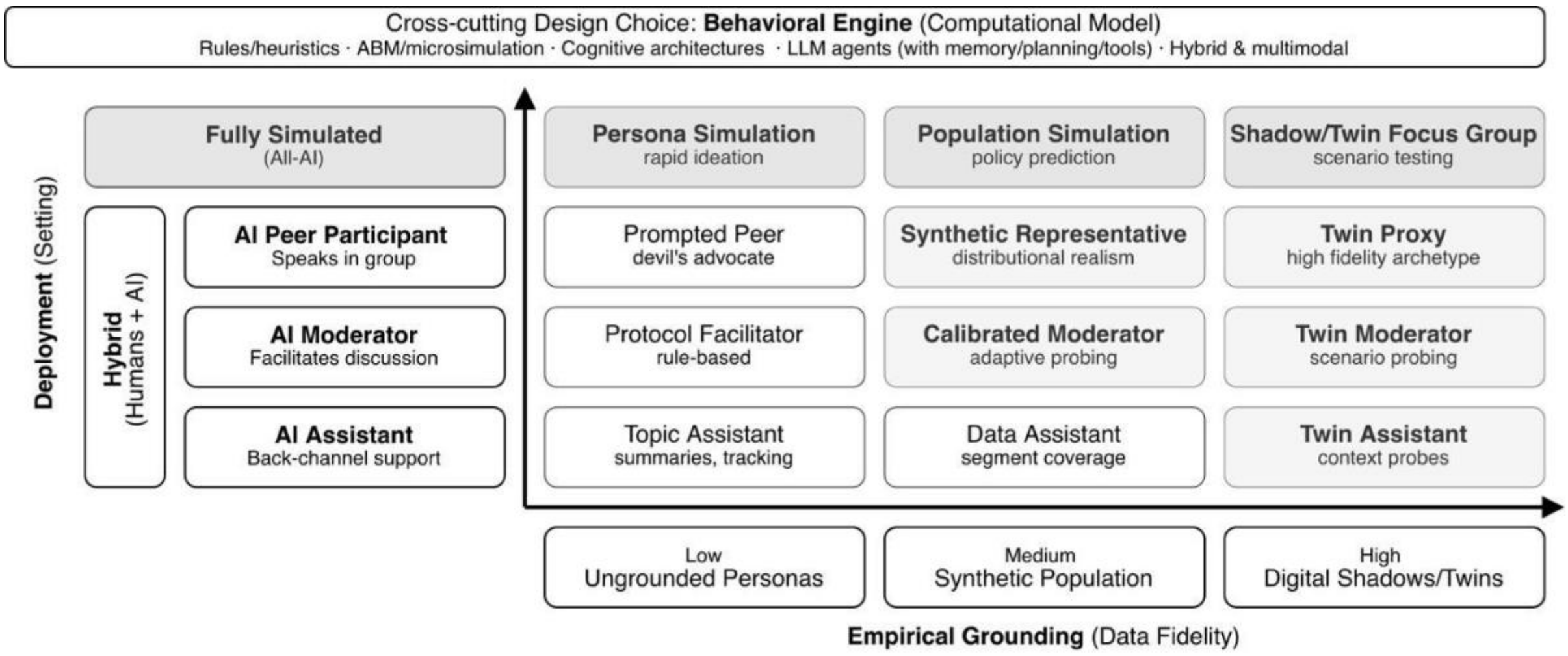

*Figure 1. Design space for AI-Enhanced Simulation Technologies (AEST) in simulated focus groups. The matrix maps configurations along Empirical Grounding and Deployment. The top band denotes the cross-cutting behavioral engine (rule-based heuristics, ABM/microsimulation, LLM-based generators, or hybrids), which shapes interaction realism, variance, and characteristic failure modes across the design space.*

As we can see, empirical grounding (or data fidelity) shows how simulated participants are anchored to empirical data. At the high end are individual-grounded replicas (digital shadows or twins) built from data about specific persons or mini-publics, with bidirectional data coupling in the twin case. In the middle sit population-grounded replicas such as synthetic populations, whose participants are modeled to match statistical distributions of a target electorate on selected variables. At the low end are ungrounded generative personas, typically LLM-based agents driven primarily by prompting with only light empirical constraints.

Deployment (or setting) shows how any of these participant models is used: in a fully simulated group (all participants artificial) or in a hybrid setting where AI assists, moderates, or participates alongside real humans. In this view, an LLM is one possible behavioral engine rather than a distinct model type. LLMs can instantiate ungrounded personas or serve as the conversational layer inside more grounded agents, and other engines are also possible, including explicit cognitive architectures that model decision processes more structurally.

### 3.1. Digital Shadows and Digital Twins

Digital shadows and twins are high-fidelity computational representations of specific real-world entities, positioned along an updating continuum from one-way (shadow) to bidirectional coupling (twin). In political science applications, a twin might represent an individual voter, a composite archetype, or a collective such as a community or electorate (Novelli et al. 2026).

These models are typically data-driven: they ingest rich information—demographics, observed behavior, survey responses, digital traces (e.g., social media activity), and related signals—to approximate the entity's state and likely responses. A key feature that distinguishes a digital twin from a static digital model is its capacity for ongoing updating as new data arrive, with the aim of preserving its usefulness as a proxy over time (Kritzinger et al. 2018). For example, a political party might maintain a digital twin of a swing-voter archetype and refresh it with new polling or contextual indicators to simulate how that archetype's opinions could shift as conditions change[11]. Trade and mainstream press report that campaigns and party-aligned organizations have begun commissioning synthetic polling and simulated voter panels from commercial providers such as Aaru and Electric Twin, whose AI agents are built from census and behavioral data to stand in for real electorates (Albergotti 2024; McKown-Dawson 2026; Shipman 2025). Because most clients remain undisclosed and independent validation is scarce, such reporting should be read with caution.

### 3.2. Synthetic Populations

Synthetic populations are large collections of artificial agents constructed to match the statistical properties of a target population on selected variables. Unlike twins or shadows, they prioritize distributional realism: the goal is not a one-to-one replica but a micro-level dataset whose joint and conditional distributions resemble those observed empirically (Petit and Pachot 2025).

In a typical workflow, researchers combine census tabulations, large-scale surveys, administrative records, and voter files to generate a synthetic micro-population of individuals, each assigned a plausible combination of attributes. Construction methods range from reweighting and iterative proportional fitting to combinatorial optimization (recombining real records to match aggregates) and synthetic reconstruction (estimating a joint distribution from aggregates) (Chapuis et al. 2022).

[11] We discuss ethical and governance implications in detail in Section 6.

To simulate behavior and change over time, synthetic populations are often paired with microsimulation or agent-based modeling (ABM), which provides the behavioral "engine" that turns distributional realism into dynamic social processes (Chapuis et al. 2022). Each agent can be equipped with behavioral components such as a probabilistic vote-choice model or an opinion-update mechanism that responds to information exposure and peer influence. More advanced models may add cognitive features. A large class of such mechanisms is studied in opinion dynamics models on networks, including bounded-confidence models and their heterogeneous or adaptive variants (Bernardo et al. 2024). Agents may also be embedded in explicit network structures that represent social ties, enabling the simulation of social influence and collective dynamics in silico (Chapuis et al. 2022).

Importantly, however, ABM and microsimulation are not limited to simple probabilistic rules: For this enrichment, "cognitive social simulation" is a well-established approach, which historically integrated explicit cognitive architectures (e.g., ACT-R, Soar, CLARION) to model more realistic decision processes and their aggregate consequences (Hélie and Sun 2015). In the context of focus group simulation, this lineage has recently evolved into language-based cognitive architectures for LLM agents, which supplement conversational fluency with structured control flows, modular memory, and explicit action-selection mechanisms (Sumers et al. 2024).

When a political party uses this approach, it might, for instance, synthetically recreate the voting-age population of a district and then simulate a campaign: agents receive campaign messages (modeled as inputs), talk to each other or react based on predefined tendencies, and finally, each decides to support or oppose the policy. The outcomes can be aggregated to predict overall opinion shifts or identify segments that react strongly.

Political and campaign research sometimes relies on synthetic or semi-synthetic electorates built by combining surveys with census and administrative benchmarks through weighting, multilevel regression, and poststratification (MRP), or related microsimulation techniques, to explore "what-if" scenarios and estimate how different voter subgroups might respond under alternative assumptions (Gao et al. 2022; Pilditch and Madsen 2021).

While the synthetic population is anchored in secondary data, the agent-based and microsimulation components do not simply project these data forward. Instead, interaction rules grounded in theories of political communication and content moderation generate emergent, non-linear dynamics, including configurations of discourse, such as opinion cascades, polarization, and unexpected coalitions, which are not directly present in the input distributions (Epstein 1999). This generative use of simulation is central to our exploratory aim and, in particular, to probe how altered moderation regimes and interaction structures could reshape collective behavior under plausible but as-yet-unobserved conditions. However, from a qualitative point of view, emergence should not be mistaken for novelty. Emergence is an unanticipated outcome but derivable from simulation, from what the modeler has encoded (Bedau 1997). Differently, exploratory research prizes novelty that lies outside the model's ontology, forcing the researcher to revise their own categories. A simulation can surprise its designers, but only within the space of its own assumptions. We therefore treat secondary data as a baseline constraint on agent attributes rather than as the sole source of inferential leverage. Our analyses focus on scenario and sensitivity exploration — e.g., how conclusions change under alternative assumptions about exposure, influence, and moderation — rather than on point prediction for any single empirical case. In this sense, synthetic populations and their behavioral engines serve as structured thought experiments about campaign dynamics, not as mechanical extrapolations of existing datasets.

### 3.3. LLMs-based Simulations (or ungrounded personas)

A third approach to simulating political discussion relies on what we call ungrounded personas: simulated participants implemented as generic LLM agents, without grounding in data drawn from any real-world population. Unlike digital shadows and twins or synthetic populations[12], ungrounded personas are driven primarily by prompting — typically a brief demographic profile and a set of political predispositions

[12] One could argue that synthetic populations and LLMs agents are compatible. In fact, to some extent, a group of LLM agents constitutes a synthetic population only when the agents are instantiated and constrained to reproduce empirically observed population distributions. Otherwise, LLM agents represent just simulated respondents rather than a population.

— with little or no empirical anchor. Their behavior reflects the default conversational norms of the LLM's training data rather than observed characteristics of any actual voter group, placing them at the low end of the grounding scale (Figure 1).

In practice, each AI agent is increasingly understood not as a single prompted persona, but as an LLM embedded in an agentic workflow (i.e., a language model augmented with structured control flows and modules for memory, action selection, and retrieval or tool use) so that it can keep coherent behavior across multi-turn interaction (Wang et al. 2024; Tissaoui 2026).

Persona specifications, such as demographic profiles and political predispositions, can still condition behavior, but in state-of-the-art practice they operate increasingly as structured priors within an architecture rather than as the sole mechanism for producing realism (Sumers et al. 2024; Park et al. 2023). The key limitation remains: without grounding in real population data, the realism of these agents depends heavily on what the underlying model has learned, making them more susceptible to the failure modes we discuss in Section 4.

### 3.4. Hybrid Human–AI Simulations

A related deployment configuration is the hybrid human–AI approach, which, as the name suggests, integrates AI elements into real human group discussions, creating a mixed-participant environment. Such integration, specifically when it comes to focus groups (and potentially other forms of political deliberations), can take several forms: AI as a moderator, AI as a peer participant, or AI as an assistant. For instance, Zhang et al. (2024) illustrate an LLM-based system that acts as a moderator, facilitating the discussion: it poses questions, prompts participants, and may summarize or probe further, much as a human moderator would.

The paper also evaluates fully simulated sessions with AI participants. In the peer participant mode, one or more AI-driven agents are introduced as members of the real human group conversation. AI here can contribute with "individual" perspectives just as another human would, except their contributions are generated by an AI model following a given persona or agenda. Accordingly, a research team might insert an AI agent that consistently advocates a particular point of view (ideally a minoritarian or more controversial one) in a discussion to see how human participants respond, or to ensure that viewpoint is represented if no human in the group holds it.

Finally, in assistant mode, the AI might not directly participate in the conversation but could provide data, real-time analysis, or suggested questions to a human moderator. In all these setups, the core idea is augmentation rather than complete automation, as humans are still in the loop, but AI is embedded to enhance the process.

## 4. What Simulated Focus Groups Capture and What They Miss

Each of the techniques above has advantages for focus-group research, beginning with the most basic: creating a conversational environment. A key advantage of LLM-based simulations, for instance, is that simulated participants can be asked questions or placed in a multi-agent conversation, and behavior generation is naturalistic and plausible: the LLM produces responses in real time, one sentence at a time, reacting to prompts or other agents' statements. Accordingly, in a focus group simulation, one can instantiate multiple LLM agents – for example, five agents each with distinct voter profiles – and have them engage in a natural language discussion (and maybe moderated by either a human or an AI moderator). This configuration enables rapid, relatively inexpensive iteration over stimuli and discussion protocols, thereby supporting early-stage hypothesis generation and structured stress-testing, even when live-group resources are limited (Zhang et al. 2024).

A second set of advantages concerns how simulations can help address the limitations of live focus groups set out in Section 2. Regarding the problems of dominant voices, groupthink, and social pressure (Section 2.1), simulations can serve as calibration tools: researchers can systematically vary group composition and facilitation rules to assess how dominance, conformity, or polarization affect observable outcomes.

In digital-twin or digital-shadow approaches, the same modeled 'individuals' can be run through multiple counterfactual group settings and moderator interventions, enabling controlled comparisons of how outcomes change when interaction parameters are adjusted.

With synthetic populations, researchers can explore alternative compositions before recruitment and identify participant mixes that may increase the risk of dominance or conformity given specified assumptions about interaction dynamics. With LLM-driven simulations, researchers can test moderation protocols (e.g., no-interruption norms or equal-airtime rules) and observe how discussion outputs change (Ashkinaze et al. 2025). Also, they can explicitly instantiate dominance-prone (or, conversely, deference-prone) personas (e.g., an agent instructed to take more turns, use more assertive language, or steer topics) and analyze the resulting transcript features.

This is not to claim that simulations facilitate social dynamics, but that they can support more informed methodological choices: diagnosing potential distortions and evaluating mitigation strategies without consuming scarce session time or exposing human participants to unnecessary risk. The point cuts deeper into party politics, where the target is often private preferences of the kind revealed in the voting booth, despite the public pressures of group discussion. In that sense, simulations can help probe the gap between privately held attitudes and publicly expressed positions under social influence.

Similar considerations can be applied to the replicability and generalizability issues (Section 2.2) of real-life focus groups. Simulated focus groups can enhance what we call procedural replicability by fixing the topic guide, moderator behavior, turn-taking rules, and group composition and then running the session repeatedly to see how stable themes are under identical procedures.[13] Researchers can then intentionally vary one element at a time (e.g., different moderator style) and measure sensitivity. That's especially valuable because, as we pointed out, in real focus groups the "data-generating environment" changes even when the guide remains the same (scandals, leadership changes, etc.). Replicability can be enhanced with different simulation models, but the best way is probably through digital shadows (or twins) as the same "individuals" across many procedural variants. Synthetic populations can also help here by rerunning the same synthetic electorate repeatedly and isolating which assumptions drive results.

In terms of what we call analytical replicability, simulations generate repeatable corpora (transcripts under known settings), which can be used for testing whether a coding scheme or interpretations yield consistent patterns across cases. Researchers can thus explicitly test: "Does our analytic frame recover the same narrative patterns when we vary context/composition?" Consistent with the qualitative character of the method, these gains should be read in qualitative terms such as dependability (Lincoln and Guba 1985) (e.g., supporting hypothesis generation and analytic consistency) rather than as importing quantitative replication standards into focus-group research .

[13] This type of approach was what was discussed originally in Novelli et al. 2026.

The remaining trustworthiness criteria do not transfer even in this limited way. Whereas repeatable corpora can support the dependability of coding procedures, the credibility of findings rests on thick description, the researcher's declared positionality, and reflexivity about how interpretive choices shape results (Lincoln and Guba 1985; Tracy 2010). Precisely, simulated sessions cannot satisfy these criteria internally: a synthetic transcript carries no lived context from which thick description could be built, and a model has no standpoint to declare. These requirements do not disappear under simulation; they migrate to the humans around it and simultaneously set the limits of the tools. Positionality attaches to those who specify personas, interaction rules, and grounding data, since every such choice encodes assumptions about which voters and behaviours matter. For instance, we situate these design choices within our positionality as researchers working in Western democratic contexts with prior empirical work on content moderation and far-right mobilisation, acknowledging how this standpoint shapes which mechanisms and failure modes we prioritise. Reflexivity attaches to the analysts who interpret simulated output, who must interrogate how far apparent findings are artefacts of specification. This migration is why our framework treats interpretive competence as a capacity the tool presupposes rather than provides, and why the procedural safeguards proposed in Section 6 should be read as reflexivity rendered institutional (Pillow 2003).

To enhance the generalizability of focus groups, simulation can be used to explore whether your interpretive takeaways (the themes and group dynamics you identified) depend heavily on the exact mix of people you happened to recruit by running the same discussion with, for instance, different social-status distributions or partisan mixes. Synthetic populations based on demographic/political distributions are probably the best simulation models for enhancing generalizability, as they can address recruitment and segmentation concerns directly: you can model how different recruitment skews might affect what gets voiced.

Finally, simulated focus groups can help address logistical and operational constraints (Section 2.3). In terms of speed, flexibility, and cost, LLM-based agent simulations are particularly well suited to these challenges, as they can be deployed rapidly to pilot and refine discussion guides and to explore multiple iterations at relatively low marginal cost. They also make it feasible to explore multiple iterations in rapid succession, including simulating successive “waves” of discussion as political contexts evolve, for example, by adapting prompts to reflect events before or after a major scandal. Hybrid human–AI groups are more expensive but can still improve the efficiency of the focus group (or its fine-tuning through repetitive iterations), for instance, by automating transcripts, providing fast summaries, finding a common deliberative ground (if the context requires it) (Tessler et al. 2024), or standardizing moderation. Synthetic populations are most helpful earlier in the research process, when researchers or practitioners must decide how to allocate limited resources across groups. In this role, they can optimize coverage by identifying which demographic segments are most likely to yield the greatest marginal insight within time (and budget) constraints. Digital shadows and twins are very powerful but very data-heavy and potentially expensive (besides being politically controversial for creating surrogates of actual people) (Fontes et al. 2024; Popa et al. 2021). That cost is easiest to justify when you're doing counterfactual analysis, because the whole point is to explore scenarios you cannot safely or cheaply test in the real world (e.g., equipment failures or new policies, redesigns) and to trace their ripple effects through a complex system. By contrast, day-to-day operational logistics typically demands fast, reliable answers under time pressure, and those decisions can often be handled well with lighter-weight forecasting and optimization methods, making a full twin/shadow unnecessary overhead for routine operations (Taillandier et al. 2025).

### 4.1. Limits of simulation technologies (for focus groups)

Avoiding a techno-solutionist perspective requires careful attention to the growing evidence on the limits of simulating social dynamics such as those found in focus group interactions. Empirical work comparing AI support in virtual focus groups to human-only moderation has started reporting potential efficiency gains alongside risks of reduced engagement or loss of nuance when AI outputs are not overseen by humans (Chen et al. 2025).

#### 4.1.1. Digital twins and shadows

In the development of digital shadows and, more critically, digital twins, the primary challenge lies in balancing data collection feasibility, privacy, and simulation fidelity. Replicating the behavior of social groups requires massive datasets, often making these models (currently) prohibitively expensive, especially since a high-fidelity twin requires a bi-directional, real-time data link.[14] Unlike physical systems with automated sensor telemetry, social data is fragmented and unstructured, requiring labor-intensive cleaning and normalization. This raises ethical concerns about access, as campaign tools built on these high-cost methods may be affordable only to well-resourced parties in high-resource countries, while underfunded parties elsewhere are priced out, as already shown in (Kreiss 2016). Furthermore, unbalanced data availability across socio-demographic cohorts risks systematic underrepresentation, a problem compounded by escalating privacy constraints.

But even with access to exhaustive, legally obtained datasets, predicting human behavior remains constrained by the limited predictability of complex dynamic systems (shaped by randomness, feedback effects, and computational constraints) and emergent phenomena (Helbing and Argota Sánchez-Vaquerizo 2023). Social interactions are shaped by non-linear dynamics, emergent patterns, and situational variability, all of which are hard to predict or reduce to formulaic representations. Digital twins risk oversimplifying reality by prioritizing quantifiable metrics over latent, qualitative variables or by neglecting more spontaneous, less goal-directed behaviors that are difficult to formalize but central to social interaction (Helbing and Argota Sánchez-Vaquerizo 2023, 81). This limitation is particularly acute in focus-group settings, where real-time, social dynamics — such as peer influence, empathy, and group chemistry — are emergent properties that historical data alone cannot predict. While shadows and twins might replicate a participant's general opinions, they may not capture how that same person might respond to a specific focus group atmosphere or a particularly charismatic moderator.

A further limitation, especially relevant for LLM-based simulation, concerns the modality gap between how current AI systems learn and how humans acquire social judgments. Most generative modeling is trained primarily on text — a narrow, lossy representation of the social world. "World models" aim at more grounded representations from rich sensory input (LeCun 2022). In focus groups, this matters because nonverbal cues, gestures, tones, and interruptions shape what is said and how, and these features are difficult to represent in text-only simulations. However, these features are difficult to represent in text-only simulations.[15]

[14] Cost considerations for digital shadows and twins fall into three layers. First, data collection has costs regardless of modeling: relevant social data may require labor-intensive manual work or ethically fraught data mining, and it remains unclear what minimum dataset (both in size and privacy sensitivity) is enough for a useful model. Second, the model layer can be expensive, especially for complex or multi-agent systems built on large language models (e.g., one per simulated persona). These demand substantial compute, but are increasingly attainable at near-consumer scale and shouldn't be conflated with truly prohibitive historical costs (like 1960s mainframes). Third, and most distinctively, the digital twin layer, that is, bidirectional, real-time data exchange, is critical: sustaining both the technical infrastructure and participants' ongoing cognitive engagement in hybrid human–AI systems is resource-intensive and still largely novel.

4.1.2. Synthetic population

Data-related challenges also affect the use of synthetic population models in social simulations. Relevant information is often difficult to obtain or may fail to adequately represent the target population (Bigi et al. 2024) — particularly for scarce data types, such as psychological attributes. This is especially problematic for hard-to-reach populations. As a result, models often get the broad strokes right yet flounder on fine-grained dynamics: e.g., a synthetic population might match a city's age and income distribution, but if it omits the social network structure (who knows whom), it cannot accurately simulate how information or influence spreads in a community. Such simplifications can be fatal for group-based interactions: a focus group in a simulation might be composed of agents with no meaningful ties or history, whereas real focus-group participants' interactions often depend on trust, status, or common ground; their views can shift due to peer pressure or alliance formation during the discussion. Several of these dynamics have been operationalized in state-of-the-art computational social simulation, although end-to-end focus group models that jointly capture trust, status, and emergent interactional dynamics remain comparatively underdeveloped (Chuang et al. 2024; Jiang et al. 2024).

Even when data are available, they can be inconsistent: the same variables may appear in incompatible formats (e.g., age recorded as an integer in one dataset and as a range in another) or be defined at different spatial or temporal scales (Chapuis et al. 2022, 5). Addressing these inconsistencies often requires bespoke synthesis tailored to specific analytical needs (Snoke et al. 2018, 664), a process that is both technically demanding and resource-intensive.

Beyond data quality, synthetic population modeling also faces methodological difficulties. A key issue is the assumption of statistical independence among attributes (common in synthetic reconstruction approaches) (Chapuis et al. 2022), when real-world relationships among variables are unknown or interdependent. Also, it is often challenging to identify appropriate validation indicators to assess how well the synthetic population reflects the characteristics of the real target group after it is generated.[16]

[15] This limitation is increasingly being addressed by multimodal large language models that integrate vision (and sometimes audio/video) encoders with LLMs, and by emerging work on multimodal social-interaction modeling that aligns verbal and nonverbal cues in multi-party settings (Yin et al. 2024). However, these advances still face substantial challenges in faithfully reproducing emergent group dynamics at scale. As a result, LLM-based simulations might be better suited to primarily text-mediated settings (e.g. forums, chats, comment threads) than to full, in-live political talk.

4.1.3. LLMs simulations (ungrounded persona)

The use of large language models (LLMs) for social simulation is growing in popularity, and with it, scholarly awareness of the method's limitations (Qu and Wang 2024). Some of these challenges are well-known and stem from general performance issues associated with LLMs, such as hallucinations (Alansari and Luqman 2025), sycophantic tendencies (Malmqvist 2025), or more recently identified degradation patterns like 'LLM brain rot' (Xing et al. 2025). Among these, sycophancy is especially relevant in simulated focus groups, as it may lead models to overproduce moderate, agreeable viewpoints, thereby generating artificial consensus and underrepresenting conflict or dissent. This bias toward reasonable output reflects a learned preference for avoiding controversial or extreme positions, and it can also pull agent behavior back toward the model's default conversational norms in multi-turn settings.[17]

There are, however, other specific issues that are even more relevant for social simulation and, therefore, for focus groups. Questions arise about AI-character consistency: do agents maintain their assigned persona throughout multi-turn discussions, or do they drift toward model defaults? A key finding, for instance, is that LLM agents often stick to the biases or norms of their training data, and that additional guidelines and guardrails are then overlaid on the model before release, even when you assign them roles that should diverge from those norms (e.g., through persona prompts). Taubenfeld et al. (2024), for example, found that in simulated political debates, GPT-based agents assigned to represent opposing ideological camps still conformed to the model's default social norms, resulting in interactions that deviated from the dynamics typically observed in real partisan discourse.

More broadly, this can yield an 'average persona' collapse: a tendency for LLMs to produce homogeneous behavioral patterns. Because model training emphasizes high-frequency tokens and mainstream language use, rare or minority behaviors are underrepresented, leading to the suppression of subcultural or divergent traits. This issue has led researchers to call for a systematic analysis of behavioral variance in LLM outputs prior to deploying them in scientific simulations (Wu et al. 2025).

[16] The size of the synthetic population compounds all these accuracy problems (Saadi et al. 2018).

[17] Also, some mitigation strategy can be found in (Wei et al 2023), including bias prompting for getting fragmentation.

Efforts to improve realism by shaping interactional behavior through persona-prompting (which can be seen as a form of vibe finetuning) have shown limits. For instance, a recent work suggests that assigning an expert identity to an LLM in a specific domain does not necessarily enhance the factual accuracy of its outputs, though it may affect stylistic features such as tone or confidence (Basil et al. 2025).

These findings raise a measurement question that any responsible deployment must confront: how to establish that persona specifications (demographic and psychographic inputs) achieve realism rather than mere projection, given that such inputs operate as causal levers within the model rather than as authentic representations of reality? The emerging answer is benchmarking for "algorithmic fidelity", that is, comparing simulated outputs against held-out human data (like survey distributions, real transcripts, or parallel human sessions) at the aggregate level (Argyle et al. 2023; Bisbee et al. 2024). No accepted accuracy thresholds yet exist that would license treating a simulation as an authentic replacement for human focus groups. In our view this absence is itself decisive: it is why the matrix (Section 5) never licenses replacement, only bounded auxiliary use with grounding requirements that escalate with risk.

Early evidence from studies applying LLMs to focus group simulation supports many of these concerns. Accordingly, the experimental paper we mentioned earlier, Zhang et al. (2024), observed that while AI-generated participants can simulate diverse perspectives, their contributions often reflect generalized or stereotypical views and lack the depth and idiosyncrasy of human responses. Moreover, across repeated sessions, simulated conversations tend to exhibit high redundancy: opinion statements are frequently repeated, and the pool of unique insights plateaus quickly, indicating limited iterative creativity (Zhang et al. 2024, 7–9).

#### 4.1.4. Hybrid Models

To conclude on the limits of the hybrid-AI simulation model, one empirical finding is that introducing AI agents into human groups can disrupt natural interaction patterns. For example, in a recent collective-creativity study where humans, AIs, and mixed teams co-wrote stories, AI-only groups initially produced more diverse/creative outputs than mixed groups because humans tried to preserve narrative continuity while AIs rewrote more freely. Mixed teams improved over time, but the early mismatch suggests humans stick to prior context, whereas AI agents, lacking memory or investment in earlier narrative, make more radical changes — so an AI participant in a focus group may feel disjointed or overly novel until people adapt (Shiiku et al. 2025). Related to this is the question of linguistic authenticity. LLM outputs may be perceived by human participants in a focus group as overly polished or formal, lacking the natural speech patterns that characterize human conversation (e.g., hedging, interruptions, filler words, and colloquialisms). Experimental evidence shows that attribution to AI can depress perceived authenticity and moral respect in emotionally meaningful communication, especially after source disclosure (Dorigoni and Giardino 2025).

When combined with the tendency to ignore conversational continuity, these linguistic and interactional differences raise doubts about whether LLM agents can faithfully simulate the group dynamics that make focus groups valuable in the first place.

A second issue concerns disclosure. AI regulations and/or industry standards normally mandate transparency around the use of AI prior to engagement. But if participants know they are interacting with AI, they may be less forthcoming (perhaps as they doubt the AI will understand nuanced political views) or may express more extreme views (as they feel freed from the social norms of speaking to a fellow human), in all cases, distorting the 'natural' flow of discussion the focus group is attempting to capture. In jurisdictions with minimal (or no) privacy and AI regulation, one approach to managing this tension would be to withhold disclosure during the session itself, while revealing AI involvement during post-session debriefings. This allows researchers to observe more naturalistic interactions while preserving informed consent and enabling participants to reflect on how the AI's presence may have shaped the discussion. However, this ex-post approach to transparency still raises serious ethical questions around the initial engagement of the human participant with the AI, especially as these focus groups are designed to elicit political and personal views of the participants, and the subsequent revocation of any consent for further processing does not in itself address any distress the individual may experience upon learning they've been interacting with an algorithm.

Evidence from focus groups with human participants suggests that AI moderators can facilitate discussion (Zhang et al. 2024), but they also risk being felt as intrusive (and dominant), looping, or repeating themselves because everyday discussion uses colloquial, fragmentary speech that makes it hard for the AI to detect whether a question has been answered (Zhang et al. 2024, 9). In focus groups, this means the design question is not just whether to disclose AI involvement, but how to structure that involvement so it helps rather than distorts the group dynamics under observation.

### 4.2. Deskilling: A Second-Order Risk

A further limit concerns what simulation adoption does to parties from the inside. When simulation outputs are ready within hours and human focus groups take weeks to organize, strategists working under electoral pressure will simply reach for the faster option. Not out of bad faith, but because time rarely favors the slower method. If this becomes routine across campaign cycles, parties gradually lose the habit and capacity for human qualitative research: communications staff become less practiced at interpreting the messiness of real voter conversations, and tolerance for ambiguity — which is intrinsic to good qualitative work — quietly erodes (Novelli and Sandri 2025).

We anticipate three objections to this argument and address each briefly. First, a critic might note that the same worry could be pressed against every automation that replaces a slower human process with a faster machine one (e.g., calculators, spell-check, GPS navigation) and that such worries are often overblown. In some cases a distinction can be made between 'cognitive offloading' of less challenging tasks, and a less desirable 'cognitive surrender' where the user's ability to think critically is ultimately degraded (Shaw and Nave 2026).

Our reply is that the capacities at stake in qualitative research are not only procedural but interpretive: reading what a focus group is doing, distinguishing polite agreement from genuine resonance, recognizing when a silence matters. These capacities are built through sustained practice and cannot be rebuilt on demand in the compressed timescale of an electoral crisis. The point is then more general: the human-factors literature on automation has documented for decades that the skills displaced by automation atrophy precisely when they are most needed for the rare interventions only humans can perform (Bainbridge 1983). The mechanism we are invoking is Bainbridge's first: skills that go unused degrade, as distinct from the complacency-in-supervision mechanism the same literature also describes.

Second, a critic might note that parties adopting AEST do not fire their qualitative researchers; the expertise still exists somewhere in the organization. But expertise at rest is not expertise in operation: when the senior practitioners who understand qualitative evidence are no longer the people campaigns turn to by default, their judgment no longer informs decisions, even if they remain on the payroll. The capacity of our matrix at section 5 presupposes operative authority, not residual headcount.

Third, the worry applies differentially. Parties with strong existing qualitative traditions have more to lose but also more institutional ballast to resist the drift. Parties without such traditions face a different question entirely: whether AEST lowers the cost of any qualitative evidence at all, which in some contexts may be a net gain.

This second-order effect matters because it is not visible in any single campaign's output. It accumulates. A party that has substituted simulation for live research over several cycles cannot easily reverse the shift when a high-stakes moment finally demands genuine qualitative judgment, because the tacit knowledge needed to exercise that judgment has thinned (Novelli and Floridi 2026). Governance of AEST in party politics is therefore not only a question of which outputs can authorize which decisions, but also of which capacities parties must maintain to remain capable of deciding at all.

## 5. Governing Simulation Technologies in Focus Group Research: A Decision Matrix

A growing technical literature addresses how to mitigate the limitations documented above (Hu et al. 2021; Hu et al. 2025; Houde et al. 2025). We do not review these strategies in detail here because our response to the limitations is primarily governance-oriented. In short, rather than asking how simulations can be made better, we ask under what conditions and for which purposes their current outputs can be responsibly used.

Stated in practical terms, the matrix reduces to three questions that any consultant or campaign researcher can ask of a proposed use: What is the focus group for? How costly would a wrong answer be? And how firmly is the simulation anchored in real voters?

To answer that question, we propose a decision matrix (Figure 2) that draws on established governance principles rather than introducing a new architecture. Its risk layers instantiate the logic of risk-based regulation, which calibrates oversight intensity to the expected magnitude and likelihood of harm rather than applying uniform rules across cases (Black and Baldwin 2010, 2012). Its empirical grounding requirements, which rise with the stakes rather than being fixed in advance, reflect the responsive regulation principle of conditional escalation (Ayres and Braithwaite 1992).

### 5.1. How the Matrix Was Constructed

The matrix is therefore analytic and prescriptive rather than empirically validated. It was developed as an analytic typology, following established guidance on concept formation and typology construction in political science (Collier et al. 2012). The procedure involved three steps. First, a narrative synthesis of three literatures — i.e., focus-group methodology in party politics (Section 2), AI-based social simulation and its documented failure modes (Sections 3–4), and risk-based regulation and AI governance — produced an inventory of the conditions under which simulated respondents succeed and fail.

Second, a dimensional analysis of that inventory identified the three recurrent axes: strategic purpose derives from the qualitative-methods distinction between exploratory meaning-making and instrumental testing (Morgan 1996; Cyr 2019); deployment risk derives from the proportionality logic of risk-based regulation (Novelli et al. 2024; Black and Baldwin 2010); and empirical grounding derives from the validation literature's finding that anchoring in real population data is the strongest predictor of output quality (Argyle et al. 2023; Bisbee et al. 2024).

Third, the resulting cell prescriptions were stress-tested against the failure modes reviewed in Section 4 and against the campaign scenario developed in Section 5.2, and adjusted wherever they licensed uses the evidence could not support.

Before the matrix can be constructed, however, a prior question must be answered: What is the focus group actually for? The answer, as should be clear at this point, is not obvious. In party politics, focus groups are used for at least two quite different purposes, which place very different demands on any simulation technology. The first is Interaction-Based Meaning-Making (Mode 1), for discovering how political meanings, identities, and justifications are formed and negotiated through live group discussion. The second is Instrumental Decision-Support (Mode 2), a tool for rapidly testing and refining campaign messages and strategic choices under time constraints.

These two modes reflect different epistemic emphases rather than wholly distinct types of knowledge, and the distinction matters for simulation purposes. Mode 1 is primarily explanatory: it seeks to understand why and how voters think and justify as they do, by observing how views emerge and shift through live interaction. But it can also carry predictive value, since understanding how political meanings are negotiated informs judgments about how voters may respond in the future. Mode 2 is primarily predictive: it seeks to identify which message, frame, or strategy is likely to perform better under specified conditions. But it can also yield (thin) explanatory insight (e.g., revealing why a particular phrase resonates). The difference, then, is one of emphasis and depth, not of kind.

What drives the matrix's first-order filter is not that one mode is purely explanatory and the other purely predictive, but that the emergent, open-ended knowledge Mode 1 produces cannot be simulated. This is where the distinction between emergence and novelty drawn in Section 3.2 is operationalized: grounded simulations support bounded what-if refining of Mode 2 but cannot supply the open-ended discovery that defines meaning-making Mode 1. Existing AESTs fit more naturally with instrumental decision-support (Mode 2) because these uses pose bounded, comparative questions and can be embedded in workflows that explicitly test robustness rather than (assuming) realism. In this role, simulations would serve as filters for political parties, helping narrow options before costly human testing. In particular, they are well-suited to:

- (2a) variant triage (rapidly comparing many message options),
- (2b) scenario stress testing (testing how responses might shift under different political or contextual conditions),
- (2c) sensitivity analysis (systematically varying key assumptions or inputs (for example, turnout levels or candidate traits) to identify which factors most strongly shape simulated responses and to assess how robust the results are).

*Table 1. The modes of focus groups and their suitability for AEST*

| Feature | MODE 1: MEANING-MAKING | MODE 2: DECISION-SUPPORT |
|---|---|---|
| Primary Goal | Exploring how identities and justifications are negotiated. | Rapidly testing/refining messages under time constraints. |
| Epistemic Aim | Explanation: Understanding "why" and "how" voters think, feel, and behave as they do. | Prediction: Identifying which strategy (e.g., campaign message) is likely to perform better. |
| Outcome | Thick explanation of emergent social dynamics. | Instrumental optimization and filtering of options. |
| AEST Suitability | Low: LLM biases and lack of emergent human realism hinder results. | High: Well-suited for bounded, comparative triage and stress tests. |

Clarifying the strategic uses of focus groups in party strategy is a necessary first step, but it is not enough to judge when AEST is appropriate. Any responsible assessment also has to incorporate two further considerations: the specific simulation technique being deployed, and the context of use, especially the level of risk and the possibility of foreseeable harm.

The approach we take, then, is to situate any prospective AEST application along three ordered axes. The first is the strategic mode already introduced: whether the focus group is being used for Instrumental Decision-Support (Mode 2) — including both early-stage hypothesis generation and more targeted message testing — or for Interaction-Based Meaning-Making (Mode 1). These represent the thin and thick ends of the epistemic spectrum, respectively, and the distinction between them functions as the matrix's first-order filter.

Second, we layer in risk (noted with R), understood as the expected downside if the output is wrong or misleading:

- R0 Sandbox: internal learning only
- R1 Operational: results inform real (strategic) choices of the party but remain limited in reach and relatively reversible (e.g., scheduling, resource allocation, drafting FAQ cards)
- R2 High-stakes: errors can propagate widely and the costs of miscalibration are substantial (e.g., paid media, targeting, and field experiments)

Finally, we differentiate AEST by their empirical grounding (or fidelity, noted with G) – as we illustrated in Figure 1 – and ranging from techniques that are least constrained by participant input to those most tightly anchored in reliable data and, even more importantly, human interaction (hybrid models with humans in the loop):

- G0: LLM-only personas (prompted role-play with no participant data)
- G1: Data-anchored synthetic participants: simulations grounded in participant data (e.g., anchored LLM sims, synthetic populations, digital shadows/twins)
- G2: Human-calibrated synthetic: any G0–G1 method plus a rapid human elicitation/calibration loop
- G3: Mixed-group: real humans in-session with some AI-simulated participants and/or AI moderation
- G4: Human focus group with AEST augmentation: a fully human group, with AI as copilot (augmentation rather than substitute participation)

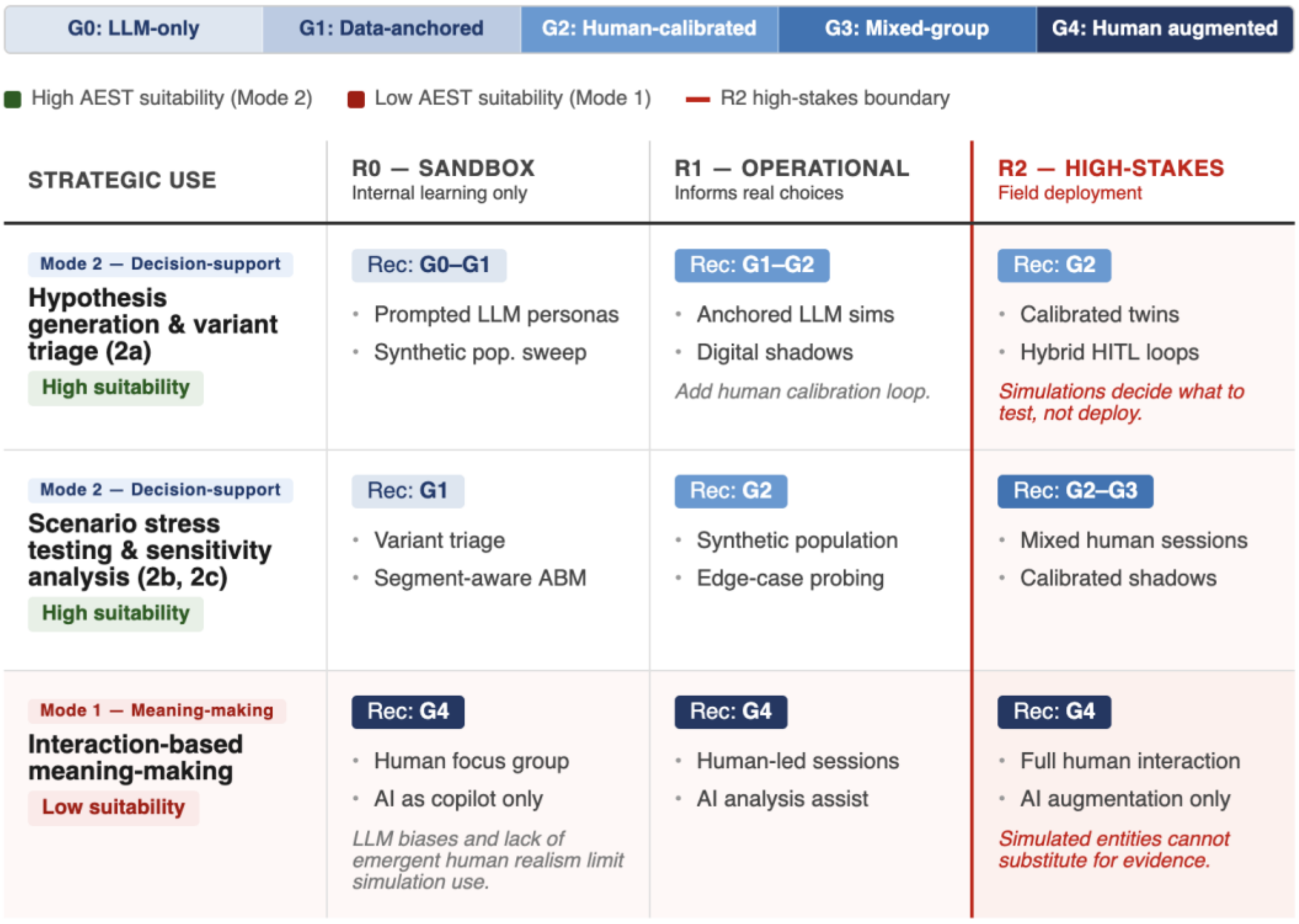

*Figure 2. AEST Strategic Framework: Grounding vs. Risk. Recommended minimum (Rec.) G-level by strategic use and risk layer. Mode 1 (meaning-making) has low AEST suitability throughout; Mode 2 (decision-support) is the primary domain of responsible AEST use. Note: "High suitability" to AESTs (Mode 2, low risk, calibrated grounding) reflects operational efficiency, not deliberative sufficiency. These outputs should be treated strictly as provisional inputs. Political judgment, ethical reasoning, and emergent consensus remain exclusively human domains. Automation drift, i.e., assuming that procedural optimization equals final evidence, must be actively resisted.*

The decision matrix shows that AEST works best for variant triage and scenario stress testing, and therefore, for exploring outcomes within a bounded design space (messages, frames, moderator rules, group composition). Party strategists can use them to rapidly test multiple message formulations before committing resources to live groups. They are less suitable and potentially dangerous unless strongly calibrated against real data or through interaction with real focus groups (G4 and G3) to capture authentic group dynamics or explore unanticipated concerns. In other words, simulations can support bounded exploratory work (e.g., hypothesis generation and sensitivity testing under explicit assumptions), but they seem to be poor substitutes for exploratory focus groups aimed at discovering what matters to participants in the first place.

The matrix also makes clear that suitability depends on both risk and empirical grounding. In R0 (sandbox), lower-grounded tools (G0–G1, such as prompted or anchored LLM personas) can be used to generate hypotheses and rough shortlists. In R1 (operational), parties should rely on better-grounded or hybrid workflows (G1–G2), using synthetic populations or shadows/twins where available and adding a human calibration loop to correct assumptions. In R2 (high-stakes), the matrix requires hybrid calibration as a minimum (G2 mandatory): simulations can inform what to test next, but not what to deploy, and any deployment decision must be backed by real empirical validation (human groups and/or experiments).

Finally, for interaction-based meaning-making (e.g., where the goal is to observe emergent negotiation of identity, justification, and political meanings), the matrix draws a hard boundary: this requires human-led focus groups (G4) at every risk level. AEST can assist with guide design, probing, and analysis, but LLM simulations, synthetic populations, and shadows/twins cannot substitute for live interaction evidence.

If the goal is explanatory in the sense of mechanism probing (e.g., how conformity pressures or message framing might shift expressed views under specified interaction rules), simulations can be useful as "what-if" tools, but conclusions must be framed as conditional on model assumptions. In stable political environments, simulations calibrated to past elections may offer useful approximations. In volatile contexts (e.g., leadership changes, scandals, major policy shifts), simulations often become unreliable because they cannot adapt quickly enough to shifting norms. In such cases, if volatility is high or the task turns out to be exploratory (new issues emerging), you loop back from simulation to live groups and then re-scope what simulations are allowed to do.

### 5.2. A Scenario: Hybrid Focus Groups on a National Election Campaign

To illustrate how the decision matrix works in practice, consider a political party preparing for a national election. The party wants to test three campaign messages (A, B, C) targeting swing voters, but faces tight deadlines and limited research budgets. Rather than running multiple rounds of live focus groups from the outset, the campaign adopts a hybrid approach that moves deliberately through the matrix.

In the initial phase, the team conducts variant triage — a Mode 2 task — using LLM agents calibrated to swing-voter demographics and past survey responses (G1 grounding). Because outputs are used only to narrow the field internally, the risk level is R0 (sandbox). Each simulation tests one message. Outputs reveal that Message A produces the most engagement but raises unexpected credibility concerns; Message B ranks second; Message C generates little interest. The simulations do their job: they filter options cheaply before any resources are committed.

In the validation phase, the risk level rises to R1 (operational): findings will directly shape which messages receive further investment. At this level, G1 grounding is no longer sufficient, and the team runs live focus groups: three sessions with real swing voters (G4). Message A performs well, but participants raise credibility concerns that differ from those flagged by the simulation, confirming that the agents missed important contextual cues. Message B elicits the strongest emotional response, described by participants as more relatable and grounded in everyday concerns. The campaign prioritizes Message B, while noting its potential vulnerability to attack.

In the stress-testing phase, the team returns to simulation for scenario analysis (again a bounded Mode 2 task) but now at R1 risk, requiring at least G1–G2 grounding. They simulate focus groups in which an opponent attacks Message B on cost, feasibility, and consistency with the party's record. The simulated groups expose vulnerabilities around cost. The campaign refines the message accordingly and re-tests it in a final live group, which confirms the revised framing is more defensible.

In the final phase, the party moves to R2 (high-stakes deployment: paid media and field targeting). At this level, the matrix is unambiguous: simulation outputs alone cannot authorize deployment. Live group validation, already completed, provides the necessary empirical anchor. Simulation insights inform the anticipation of counterarguments; live group findings finalize tone and framing.

The transitions between phases are driven by the matrix logic: as risk increases, grounding requirements rise, and simulation progressively gives way to live validation.

Neither method alone would have produced the same result, but the matrix is what structures their interaction. This workflow models a transitional protocol, not a replacement pathway. The party's research team may operate under a strict internal governance rule: simulation narrows options, but only human deliberation authorizes them. The matrix is designed to prevent automation drift (not enable it).

Something to stress here is that the scenario above assumes that the party's research team operates with a shared understanding of what simulated evidence can and cannot authorize. In practice, as mentioned above, as political parties are not monolithic structures, scenario outputs will be interpreted by multiple internal audiences, such as campaign directors, communications staff, and elected representatives, each of whom may read the same simulation differently depending on their factional position or policy priorities. The hybrid protocol, therefore, requires an internal literacy program that prepares these audiences to evaluate and contest simulated evidence on its own terms.

The matrix presupposes analysts equipped with qualitative literacy, including the reflexivity that classical focus-group interpretation demands (Berger 2015), and its prescriptions are only as good as the interpretive competence applied to them. Different analysts will not read simulated data identically; what the classical exploratory approach supplies through researcher reflexivity must here be supplied procedurally. Formalizing this competence as a procedural extension of the framework is a priority for further research.

## 6. Ethical and Governance Considerations

Deploying AESTs in focus-group research raises ethical issues that are not separate from our argument but are part of its foundation. They help explain why the decision matrix proposed in this paper takes the form it does. The matrix goes beyond being a methodological guide. It is also a way of governing ethical risk. Its three core dimensions — strategic purpose, deployment risk, and empirical grounding — reflect recurring concerns about consent, privacy, transparency, accountability, and misuse. These concerns correspond to the principles that recur across the global corpus of AI-ethics guidance — transparency, justice and fairness, non-maleficence, responsibility, and privacy (Khan et al. 2022; Floridi et al. 2018; Novelli et al. 2024). Each maps onto a specific operation of the matrix: consent and privacy constrain the grounding dimension (which participant data may be used to construct simulations), transparency governs reporting obligations at every risk level, and accountability scales with the stakes dimension (who must sign off, and on what documentation, as risk rises).

The first concern is informed consent. In particular, in hybrid human-AI focus groups, participants should know in advance whether AI agents or AI moderation are involved, and their consent should be obtained accordingly. In research designed to elicit political opinions or personal views, non-disclosure should be treated as presumptively unacceptable. The fact that disclosure may influence behavior is not, by itself, a good enough reason to conceal AI involvement. If delayed disclosure were ever considered, it should be exceptional, tightly justified, independently reviewed, and followed by immediate debriefing and a meaningful opportunity to stop further data processing. This helps explain one of the matrix's clearest boundaries: where the purpose of the focus group is to observe meaning-making through live interaction, ethical and methodological concerns converge. The case against substituting AI for human interaction is not only epistemic but also ethical.

A second concern is privacy. In some cases, AEST may reduce the need to collect fresh political opinions directly from participants. But more highly grounded systems — such as synthetic populations, digital shadows, or digital twins — depend on personal data at the point of construction. In political contexts, this may involve especially sensitive information, including data that reveals political opinions. That creates strong obligations: explicit consent where required, data minimization, anonymization or pseudonymization, and the use of privacy-enhancing techniques. This is one reason the decision matrix does not simply treat more grounded systems as better. Greater grounding may improve usefulness, but it can also increase privacy burdens. The matrix is therefore structured around trade-offs: as uses become riskier, stronger grounding may be needed, but so are stronger safeguards.

Transparency (and reliability) is equally important. Simulated focus groups do not produce voter opinions; they produce computational approximations under specific assumptions. For that reason, parties should clearly distinguish simulated outputs from live evidence in internal reporting, and they should never present simulated findings externally as if they were authentic public opinion.

Focus groups have always been simulations of a kind: Gould's and Morris's swing-voter panels were staged approximations of the electorate, and their findings acquired an authority inside New Labor and the Clinton White House that their evidential base could not support (Morris 1997; Gould 1999). AEST compounds this problem with a new pseudo-credibility, while the researcher accountable for the underlying assumptions becomes harder to identify. A manufactured consensus that once required a rented room and a persuasive memo can now be generated at scale and presented with the veneer of science.

This concern also helps justify the matrix's structure. In low-risk, sandbox settings, simulations may be used for exploratory purposes. But as one moves toward operational or high-stakes use, the danger of treating simulated output as genuine evidence becomes much more serious. That is why the matrix requires more grounding, more human calibration, and ultimately live validation as the stakes rise.

Accountability introduces a further layer. AESTs are attractive partly because they are fast, scalable, and cheap to iterate. Parties can test many messages and policy variants with far less friction than human focus groups require. But this convenience can change how evidence functions inside parties. Simulation can shift from a tool for testing proposals to one for selecting policy content primarily on the basis of predicted popularity. For this reason, the matrix should be paired with procedural safeguards. For example: reporting documentation of simulation design and reasoning, separation between the team commissioning a simulation and the team interpreting it, an internal challenge function able to request alternative simulations or human focus groups, and an external audit for high-stakes uses. These safeguards matter most at the upper end of the matrix, where the costs of error are highest.

There is one last ethical challenge that is worth mentioning, namely the use of AEST for adversarial and offensive purposes. The paper has so far assumed that parties use AEST to understand their own voters better. Yet this assumption can easily be violated. The same tools that allow a party to simulate how its own supporters respond to a policy message can be used to model the opposing party's voter base, to stress-test attack messages against synthetic personas drawn from hostile demographics, or to calibrate disinformation for specific micro-audiences. Such techniques need not be limited to local political parties, but could be of interest to, for instance, hostile nation-states. None of these applications requires any capability beyond what the paper has already described. They follow directly from the same decision matrix, applied with a different strategic objective. Nor can it be assumed that these tools will be used only by actors committed to democratic norms. A demagogic leadership (inside or outside a party) could deploy simulated "voices of the people" to legitimate positions already chosen, to manufacture evidence of grassroots support, or to justify silencing internal dissent, with no participants able to contradict the record because none ever existed.

Adversarial applications should be addressed explicitly. We propose that the following uses be treated as categorically impermissible regardless of their methodological validity: (a) simulating the voter base of another party for the purpose of designing attack communications; (b) using synthetic persona outputs to calibrate messaging intended to suppress turnout among targeted groups; (c) producing disinformation content whose framing has been optimized through simulated focus group iteration. These prohibitions are, obviously, not self-enforcing, but naming them explicitly in a governance framework creates an accountability reference point that currently does not exist in the literature or in party practice.

## 7. Conclusion

The paper has argued that AEST can extend the evidential reach of focus group research for political parties while leaving its most important epistemic weaknesses unreduced.

The decision matrix we propose offers party strategists and campaign practitioners a structured means of matching research needs to appropriate simulation technologies and calibrating empirical grounding requirements to both the strategic purpose of the focus group and the stakes of deployment. Its central lesson is that AEST suitability is sharply mode-dependent: in Mode 1 (Interaction-Based Meaning-Making), simulations cannot substitute for live human groups at any risk level, because documented failure modes such as sycophancy, persona drift, and the suppression of minority viewpoints undermine the emergent dynamics through which political meaning is negotiated. In Mode 2 (Instrumental Decision-Support), by contrast, AESTs can responsibly support bounded, comparative tasks such as variant triage, scenario stress testing, and sensitivity analysis, provided that grounding requirements rise with risk and that their outputs inform rather than authorize decisions.

What it adds to existing accounts is a warning about the second-order effects of adoption: the transformation of parties' internal decision-making structures as simulation becomes routine. Managing these effects requires treating governance as a design question and a structural guardrail: parties must institutionalize the rule that simulation may inform what is tested, but never what is ultimately endorsed, and that the capacity for human deliberation must be actively preserved through institutional governance and conscious practice, remaining the ultimate, non-algorithmic authority in the democratic process.